\documentclass[mathleft
]{an}
\usepackage{graphicx}
\usepackage{times}
\newcommand{\aap}{A\&A}
\begin{document}

\Pagespan{482}{492}
\Yearpublication{2009}%
\Yearsubmission{2009}%
\Month{11}%
\Volume{330}%
\Issue{5}%

\title{Variability of young stars: \\
        Determination of rotational periods of weak-line T Tauri stars in the Cepheus-Cassiopeia star-forming region\thanks{Based on observations obtained with telescopes of the University Observatory Jena, which is operated by the Astrophysical Institute of the Friedrich-Schiller-University.}}

\author{A. Koeltzsch\inst{1}\fnmsep\thanks{Corresponding author:
  \email{alex@astro.uni-jena.de}\newline}
\and M. Mugrauer\inst{1}
\and St. Raetz\inst{1}
\and T.O.B. Schmidt\inst{1}
\and T. Roell\inst{1}
\and T. Eisenbeiss\inst{1}
\and M.M. Hohle\inst{1,2}
\and \\ M. Va{\v n}ko\inst{1}
\and Ch. Ginski\inst{1}
\and C. Marka\inst{1}
\and M. Moualla\inst{1}
\and K. Schreyer\inst{1}
\and Ch. Broeg\inst{3}
\and R. Neuh{\"a}user\inst{1}
}
\titlerunning{Variability of young stars}

\institute{
Astrophysikalisches Institut und Universit{\"a}ts-Sternwarte Jena, Schillerg{\"a}{\ss}chen 2-3, 07745 Jena, Germany
\and
Max Planck Institute for Extraterrestrial Physics, Giessenbachstra\ss{}e, 85748 Garching, Germany
\and
Space Research and Planetary Sciences, Physikalisches Institut, University of Bern, Sidlerstra\ss{}e 5, 3012 Bern, Switzerland
}

\received{2009 Jan 21}
\accepted{2009 Apr 11}
\publonline{2009 May 30}


\abstract{We report on observation and determination of rotational periods of ten weak-line T Tauri stars in the Cepheus-Cassiopeia star-forming region. Observations were carried out with the Cassegrain-Teleskop-Kamera (CTK) at University Observatory Jena between 2007 June and 2008 May. The periods obtained range between 0.49\,d and 5.7\,d, typical for weak-line and post T Tauri stars.}

\maketitle

\section{Introduction}
T Tauri stars (TTS) are young ($\leq\,10^8$\,yr), low mass ($<3\,M_\odot$), late spectral type (G0 or later) pre-main sequence stars (PMS). Because of their youth they show high Li abundance. According to the equivalent width of the H$\alpha$ emission line ($\lambda=6563$\,\AA) TTS are distinguished between classical TTS (cTTS, $W_\lambda\geq10$\,\AA) and weak-line TTS (wTTS, $W_\lambda<10$\,\AA).\\
\indent The TTS class was defined by Joy\,(1945). TTS show photometric variability; regular variability caused by hot or cool spots spread over large areas of the surface, causing a short-period increase and decrease of flux due to the rotation of the star. Additionally they also show irregular variability, e.g. in the form of flares.\\
\indent TTS and other young objects are associated with nearby molecular clouds. These star-forming regions (SFR) are the origin of most low mass stars and massive stars.\\
\indent One such molecular cloud is the Cepheus-Cassiopeia SFR located at $l\sim100$\degr$-140$\degr and $b\sim5$\degr$-25$\degr. Part of this cloud was discovered by Hubble in 1934 and was named Cepheus Flare. First studies of the chemical composition by Lebrun\,(1986) revealed the molecular nature of the Cepheus Flare and its surroundings. Grenier et. al\,(1989) found, that the Cepheus molecular cloud is part of a giant molecular cloud complex comparable to that of Orion or Taurus and that this complex extends to the nearby constellation Cassiopeia. The total molecular mass of this cloud complex is about $10^5\,M_\odot$. Kun\,(1998) performed the first systematic study on star formation in the Cepheus Flare and found that it consists of three different layers of interstellar matter, concentrated at distances of 200\,pc, $300\pm30$\,pc and $450\pm100$\,pc, respectively.\\
\begin{figure}[ht]
\centering
\includegraphics[width=0.48\textwidth]{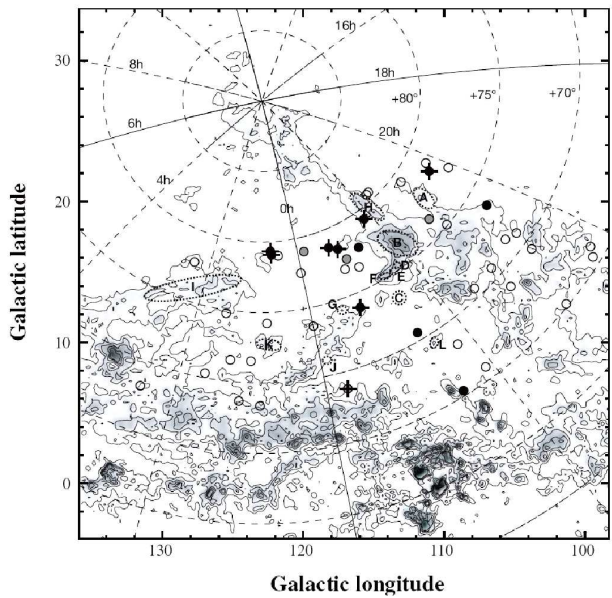}
\caption{The galactic position of the 46 previously spectroscopic observed objects. The circles (black, grey, without filling) mark the observed objects with detection of strong Li, weak Li and no Li, respectively. For further information see Tachihara et al.\,(2005).}
\label{Gal}
\end{figure}
Spectroscopic observations started in 2000. With these observations Tachihara et al.\,(2005) tried to find possible TTS. Out of 46 ROSAT selected X-ray sources 16 objects showed strong Li abundance and are a few Myr young. The galactic position of the 46 targets is given in Fig.\ref{Gal}.\\
The ten objects observed and presented here were selected from Tachihara et al. (2005).

\section{Observation and data reduction}
All observations were carried out at the University Observatory Jena which is located close to the village Gro\ss{}schwabhausen west of Jena. The monitoring of the different fields in Cep-Cas-SFR was carried out with the Cassegrain-Teles\-kop-Kamera (CTK). The CTK is installed at the 25\,cm auxilary Cassegrain-telescope which is mounted on the tube of the 90\,cm mirror telescope of the observatory. The characteristics of the instrument are described in more detail in Mugrauer (2009). The characteristics of the CTK are listed in Table \ref{CTK}.
\begin{table}[ht]
\centering
\caption{Characteristics of the CTK}
\label{CTK}
\begin{tabular}{lr} \hline 
Detector: & CCD TK\,1024 (Tektronix)\\
Pixel: & 1024\,x\,1024 (24\,$\mu$m)\\ 
Pixelscale: & (2.2065 $\pm$ 0.0008)''/Pixel \\
Field of View: & 37.7' x 37.7' \\
Filter: & B, V, R, I (Bessel),Gunn z \\ \hline
\end{tabular}
\end{table}
\subsection{Target selection}
Our target selection is based on the puplication by Tachihara et al.\,(2005): 16 stars out of 46 were found to be TTS.\\ Out of this sample ten objects were choosen. Candidates for binaries were excluded (except 36\,c1, see Table \ref{GSHobs}) because they were visually unresolvable with the CTK or the obtainable accuracy to resolve the binary candidates would have been too low. Table \ref{GSHobs} lists the observed objects.
\begin{table}[ht]
\centering
\caption {Observed objects. All data (including spectral type and age) were taken from Tachihara et al. (2005)}
\label{GSHobs}
\begin{tabular}{ccccc}\hline
GSC & No & V & Spectral & Age\\
name & & [mag] & type & [Myr]\\ \hline
0458901101 & 19 & 10.39 & G8 & 6\\
0445900083 & 20 & 10.62 & K0 & 4\\
0458601090 & 28 & 11.66 & G8 & 25\\
0460801986 & 34 & 13.09 & K7 & 3\\
0427200174 & 36\,c1 & 12.92 & K4 & 20\\
 & 36\,c2 & 14.69 & M4 & 0.2\\
0460400743 & 38 & 11.88 & K2 & 8\\
0460502733 & 40 & 10.98 & K0 & 6\\
0460500170 & 44 & 11.77 & K4 & 2\\
0447900348 & 45 & 12.64 & K2 & 25\\
0461001318 & 46 & 11.34 & K1 & 6\\ \hline
\end{tabular}
\end{table}
Because the constellations Cepheus and Cassiopeia are located on the northern hemisphere at declinations between 65\degr and 79\degr, we could observe them all year. Observations were started in July 2007 and finished in May 2008. The filters used are Bessel $V$, $R$ and $I$. During the observation we took 3-5 skyflats with an exposure time of 10\,s within every filter during dusk or dawn. Additionally we took dark exposures for flatfield and science images. See Table \ref{GSHtime} for the observation log.

\subsection{Data reduction}
\begin{figure}[ht]
\centering
\includegraphics[width=0.48\textwidth]{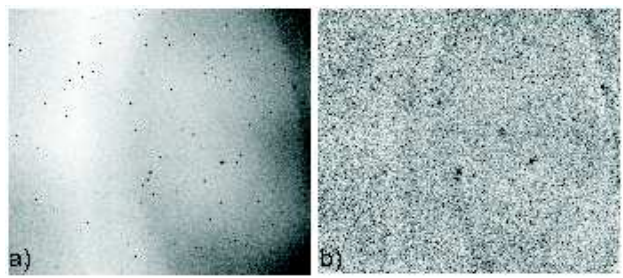}
\caption{The same image a) before (left) and b) after illumination correction (right). The image was taken on December 21, full moon was December 24. The gradients caused by sharp blendings between dark and bright areas could not yet be removed.}
\label{ic}
\end{figure}

\begin{figure}[ht]
\centering
\includegraphics[width=0.48\textwidth,height=0.22\textheight]{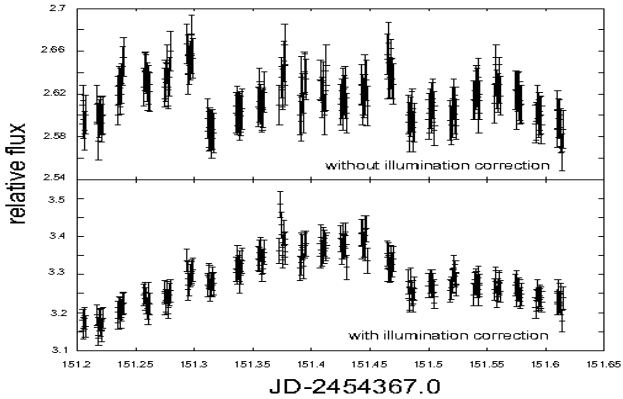}
\caption{The light curves without and with illumination correction. The data belong to the observation of object 19 on December 22.}
\label{lc_ic}
\end{figure}

Data reduction was done using the standard IRAF\footnote{IRAF is distributed by the National Optical Astronomy Observatories, which are operated by the Association of Universities for Research in Astronomy, Inc., under cooperative agreement with the National Science Foundation.} procedures \textit{flatcombine}, \textit{darkcombine} and \textit{ccdproc}. In addition, most of the images had to be corrected by means of an illumination correction. For this, the IRAF task \textit{mkillumcor} was used. These corrections were necessary because of reflections on the outer area of the  tube of the 90\,cm telescope caused by the moon and nearby bright stars. Figure \ref{ic} shows the influence of the illumination correction on the image and Figure \ref{lc_ic} shows the influence on the light curve obtained after photometry.

\section{Photometry and time series analysis}
\subsection{Photometry}
The first step after data reduction is aperture photometry. Therefore we use the IRAF task \textit{chphot}\footnote{The first version of the script was written in September 2001 by Viki Joergens using the \textit{qphot} package. This script was modified by Christopher Broeg in September 2001 using the \textit{phot} package in IRAF.}, a procedure based on the standard IRAF procedure \textit{phot}. With \textit{chphot} it is possible to do photometry on every star in the field and every single frame.\\
\indent After this we do differential photometry. The problem of this type of photometry is to find an optimum comparison star (CS), which is not variable itself and has a good signal-to-noise-ratio (S/N). Broeg et al.\,(2005) developed an algorithm to create an artificial cs out of as many stars as possible. The CS is calculated as the weighted average. The individual stars are weighted according to their S/N, variability, and their presence on every image. To improve the S/N of the artificial CS, stars with low weights are removed, so that the most constant stars are used as the basis for the calculation. With this artificial comparison star the differential magnitude of the scientific object can be calculated for every single frame. To calculate the artificial CS for the objects in the Cep-Cas SFR photometry was done on all nights the object had been observed.

\subsection{Time series analysis}
To determine the rotational periods, three different methods were used: a fourier transform given by \textit{Period04}, written by Lenz \& Breger (2005); a modified version of Lomb-Scargle (LS) periodogram analysis, and a stringlength algorithm described by Dworetsky\,(1983) and compiled as a FORTRAN routine by Christopher Broeg.\\
\indent The three methods work as follows: The \textit{Period04} periodogram analysis uses a fourier transform to calculate possible frequencies (periods). A second step, to improve the obtained period, is a least square fitting, applying the fitting formula
\begin{equation}
f(t)=Z+\sum{A_i\sin(2\pi(\Omega_it+\phi_i)),}
\end{equation}
where A is the amplitude, $\Omega$ the angular frequency, $\phi$ the phase and Z the zero point. A third step to compare different periods is a calculation of uncertainties, using the error-matrix of the least-square fitting. The Lomb-Scargle periodogram analysis is equal to a least square fitting of sine waves $y=a\cos\omega t+b\sin\omega t$. The modification of the Lomb-Scargle algorithm used here is the artificial decrease of stepsize by sampling the power spectrum in a loop, using in each loop step the maximum number of frequencies still in agreement with the number of independent frequencies. This is done to get a better sampling for objects with sparse data points.\\ The stringlength algorithm is different to these two algorithms. It is normally used for \textit{``...sparse randomly spaced observations...''} (Dworetsky\,1983). Adjacent data points are folded with test periods. The sum of the square of the distance of adjacent datapoints is calculated. The sum that is minimal is the most probable period. The advantage of this method is that it works without any assumption of sine or cosine shape of the lightcurve. But as mentioned before it works best for sparse data, which is not the case in most of the data used here. In addition to this the stringlength algorithm has problems to determine short periods and determines in most cases a multiple of the correct period due to the sampling. To cope with this the sampling has to be refined for short periods. The methods were used to decide whether all algorithms produce reliable periods and if not, which method seems to be the most reliable one, respectively.
A problem occured with the modified version of the Lomb-Scargle algorithm translating power into false alarm propability (FAP). To translate power into FAP a polynom was used to fit the obtained power-FAP diagram. For high and low propabilities this polynom does not fit well. This leads to a wrong translation of power into FAP. For each obtained period the power spectrum is given. The highest power corresponds in every case to a FAP $<0.001$ but cannot be given explicitly. Nevertheless, to give an error of the period obtained with Lomb-Scargle, the generalised Lomb-Scargle (GLS) periodogram, described in Zechmeister \& K\"urster (2009), was applied. The GLS gives a correct FAP and in addition to it the $\chi^2$ according to the residuals.\\
One step to assess the quality of the determined period was to do a phasefolding with the possible periods. This visual inspection was used to determine whether an obtained period can be possible or not. In most cases one of the obtained periods could be excluded.

\section{Rotational periods}
Within the next sections the results of the periodogram analysis and the phasefolded light curves in $V$, $R$ and $I$ are given. The periods lie between 0.49 and 5.7 days. The resulting plots of the period search are listed within the appendix. Table \ref{GSHtime} lists all observed objects with their exposure times and the total number of images used within every filter. Altogether we use 6742 science exposures obtained in 77 nights.
\begin{table}[ht]
\centering
\caption {Overview of exposure times and images used for the photometry and period search of the observed objects.}
\label{GSHtime}
\begin{tabular}{cccccc}\hline
object & filter & $t_\mathrm{exp}$  & No  \\
No$^a$ & & [s] & of images \\ \hline
19 & $V$ & 30 & 528 \\
 & $R$ & 30 & 646 \\
 & $I$ & 30 & 642 \\ \hline
20 & $V$ & 40 & 203 \\
 & $R$ & 40 & 198 \\
 & $I$ & 40 & 209 \\ \hline
28 & $V$ & 30 & 121 \\
 & $R$ & 30 & 121 \\
 & $I$ & 30 & 118 \\ \hline
34 & $V$ & 60 & 65 \\
 & $R$ & 60 & 67 \\
 & $I$ & 60 & 78 \\ \hline
36c1 & $V$ & 30 & 41 \\
 & $R$ & 30 & 48 \\
 & $I$ & 30 & 50 \\ \hline
38 & $V$ & 60 & 91 \\
 & $R$ & 60 & 101 \\
 & $I$ & 60 & 90 \\ \hline
40 & $V$ & 60 & 296 \\
 & $R$ & 60 & 240 \\
 & $I$ & 60 & 221 \\ \hline
44 & $V$ & 60 & 510 \\
 & $R$ & 60 & 513 \\
 & $I$ & 60 & 508 \\ \hline
45 & $V$ & 60 & 271 \\
 & $R$ & 60 & 266 \\
 & $I$ & 60 & 275 \\ \hline
46 & $V$ & 30 & 87 \\
 & $R$ & 30 & 88 \\
 & $I$ & 30 & 82 \\ \hline
\end{tabular}
\\$^a$ Number of object taken from Tachihara et al. (2005)
\end{table}
\subsection{GSC\,0458901101 (No 19)}
\begin{figure}[ht]
\centering
\includegraphics[width=0.45\textwidth,height=0.275\textheight]{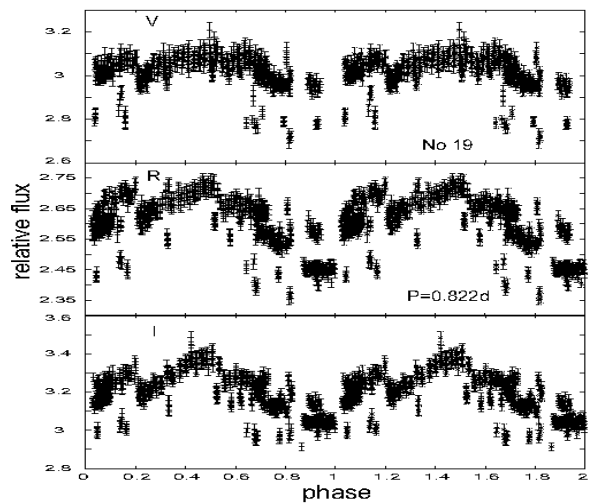}
\caption{Light curves of object 19 folded with P=0.822\,d ($V$ top, $R$ middle, $I$ bottom).}
\label{o19}
\end{figure}

\begin{figure}[ht]
\centering
\includegraphics[width=0.48\textwidth]{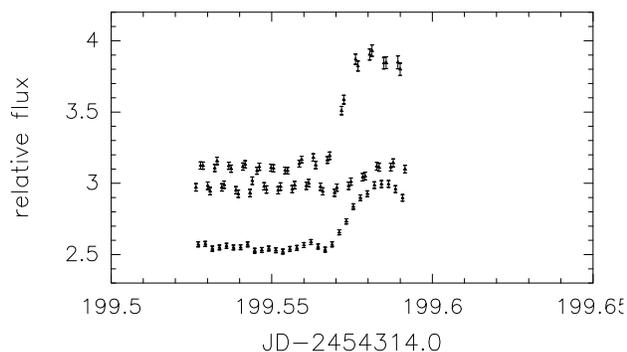}
\caption{The possible flare of object 19 from February 16 in comparison between $I$ (top), $V$ (middle) and $R$ (bottom) filter. The highest maximum is visible in $I$.}
\label{o19_fl}
\end{figure}

GSC\,0458901101 is a 6\,Myr old G8 star. The star had been observed within 32 nights, 8 of them in a series of at least one hour. To create the artificial CS only eight objects out of 576 detected were used. This small number of CS has two reasons: No star within the field seems to be very constant, and depending on the selection of the CS this object shows different light curves in $V$, $R$ and $I$. Hence, only those stars were used which are weighted higher than $10^{-2}$ within every filter.\\
\indent The period search gave no clear result. \textit{Period04} obtaines a period of 1.086\,days, the Lomb-Scargle algorithm obtaines two different periods, $P_1=0.822$\,d and $P_2=1.086$\,d\footnote{In the $I$ filter period 1 is the first maximum, in the $R$ filter the second maximum}. On the basis of the phase folding it is not possible to exclude one of these possibilities. In Fig. \ref{o19} the folding with the period 0.822\,d is shown. The stringlength algorithm obtained a period of 3.8\,d. This period could be excluded with help of the phase folding.\\
Beside this short-period variation the star showed an irregular variation in form of a possible flare. This flare was observed on 2008 February 16 (Fig. \ref{o19_fl}). The flare shows a maximum in the $I$-band with a relative flux variation of 0.8. The relative flux variation in $R$ is about 0.4 and 0.2 in $V$.

\subsection{GSC\,0445900083 (No 20)}
\begin{figure}[ht]
\centering
\includegraphics[width=0.48\textwidth,height=0.15\textheight]{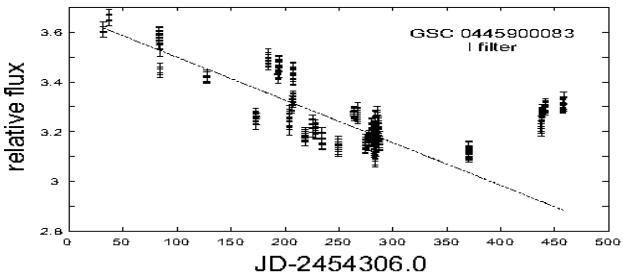}
\caption{Object 20 shows a constant decrease of flux until 2008 May. This long-term variation had to be removed to calculate the short-period variations. This was done by linear regression. The observation beginning in July show an increase of flux, but these datapoints were excluded for the calculation of the short period.}
\label{o20_long}
\end{figure}

\begin{figure}[ht]
\centering
\includegraphics[width=0.45\textwidth,height=0.275\textheight]{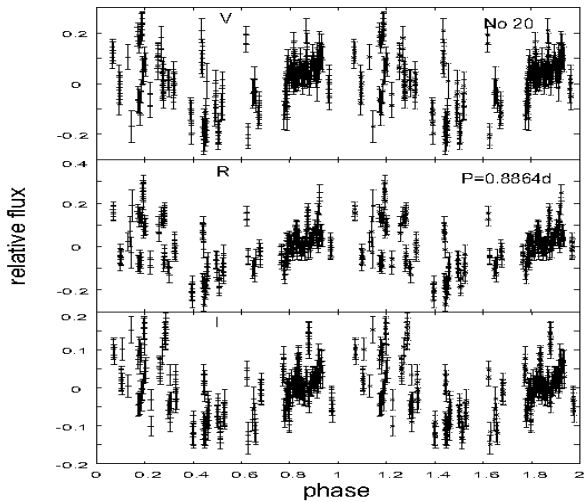}
\caption{Light curves of the residuals of object 20 folded with a period of P=0.8864\,d.}
\label{o20_res}
\end{figure}

The star GSC\,0445900083 has an estimated age of 4\,Myr and spectral type K0.\\
It had been observed within 32 nights until 2008 June 1. In 2008 July and October six additional observations were made. To create the artificial CS 25 out of 527 detected objects were used.\\
The star shows two different periodicities: a short-period variation with a period of $0.8864\pm0.0001$\,d and a long-term variation of unknown period. First period determinations with \textit{Period04} gave a maximum at $F=0.0024\,d^{-1}$. This is equal to a period of 417\,d. Because the whole observation time was only a year, this period is not reliable (see Fig. \ref{o20_long}). Further data points obtained in July and October 2008 show an increase of flux. As only one minimum and no maximum could be observed no reliable estimation of the long-term variation can be done yet.\\
A search in the residuals was used to determine the short-period variation. Lomb-Scargle and \textit{Period04} show similar results with a possible period of $P=0.886$\,d and a FAP of $10^{-18}$, the stringlength algorithm varies between 2.54\,d in $V$ band, 2.66\,d in $R$ band, and 4.77\,d in $I$ band. The folded light curves are given in Fig. \ref{o20_res}. To improve the obtained result (here we give $P=0.886\pm0.002$\,d, averaged over the results of LS and \textit{Period04}) it is necessary to determine the period of the long-term variation. 

\subsection{GSC\,0458601090 (No 28)}
The star GSC\,0458601090 is a G8 star and with an estimated age of 25\,Myr one of the oldest stars within the sample (beside GSC\,0447900348). For the analysis 27 nights were used. On the image 564 objects were detected. Out of them 30 stars were used to create the artificial CS.\\
\indent The period search algorithms resulted in two slightly different periods, $P=2.114$\,d for $R$ and $V$ filter and 2.14\,d for $I$ filter. These periods belong to the highest peaks in the \textit{Period04} and Lomb-Scargle periodograms, respectively. In both cases the other period value corresponds to the second or third strongest maximum. The Stringlength method yields three different minima within every filter but also gives a significant minimum at $P=2.14$\,d. With $10^{-28}$ the lowest FAP is obtained for the 2.13\,d period. The final period we give here is $P=2.13\pm0.01$\,d.
\begin{figure}[ht]
\centering
\includegraphics[width=0.45\textwidth,height=0.275\textheight]{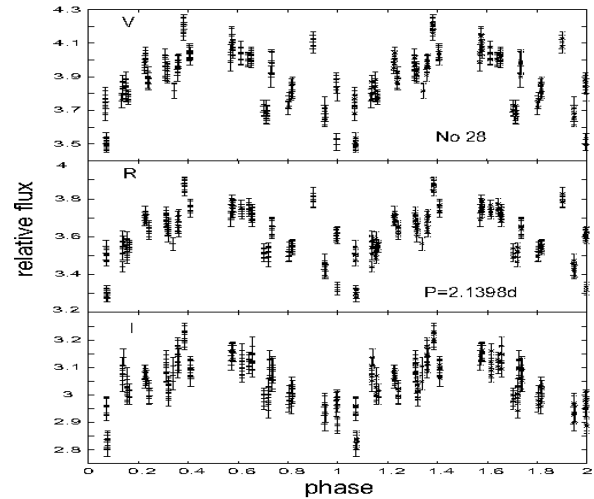}
\caption{Light curves of the object 28 folded with a period of 2.1398\,d.}
\label{o28}
\end{figure}

\subsection{GSC\,0460801986 (No 34)}
\begin{figure}[ht]
\centering
\includegraphics[width=0.45\textwidth,height=0.275\textheight]{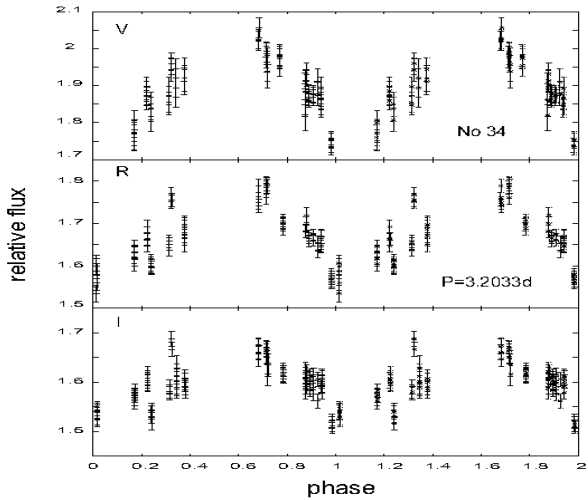}
\caption{The folded light curve of object 34. For the folding the period $P=3.2033$\,d was used.}
\label{o34}
\end{figure}

GSC\,0460801986 is a K7 star and has an estimated age of 3\,Myr. It had been observed within 26 nights, 19 of them were used for the analysis. To create the artficial CS about 20 individual stars within the field were used, 808 objects were initially detected.\\
\indent With the algorithms two different periods were found: a 1.074\,d period and a 3.204\,d period (see Fig.\ref{o34}), which is about three times the period of 1.074\,d. Examining the phase folding excludes the 1.074\,d period.

\subsection{GSC\,0427200174 (No 36)}
\begin{figure}[ht]
\centering
\includegraphics[width=0.45\textwidth,height=0.275\textheight]{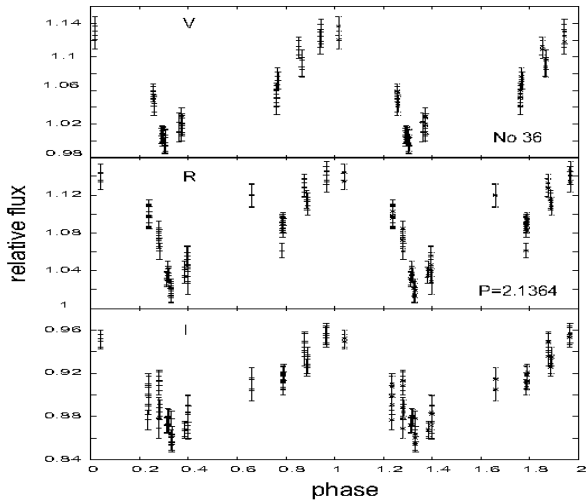}
\caption{The folded light curve of the K4 star 36\,c1, a binary-pair candidate. The folding was done with the period $P=2.1364$\,d.}
\label{o36}
\end{figure}

The object with the identifier GSC\,0427200174 is the only object in the sample that is a binary-pair candidate. It consists of an K4 star (c1) and an M4 star (c2). The determination of the period was only carried out for c1, because c2 was not visible in every image depending on the weather conditions. The star had been observed in 14 nights. To create the artificial CS about 20 stars were used, 1174 objects had been initially detected.\\
\indent Two different periods were found, a 1.873\,d period, obtained with the stringlength algorithm for the filter V and I and a 2.136\,d period, with a FAP of $10^{-18}$, obtained with Lomb-Scargle. The small amount of data points and the possibility of light of the M-star inside the aperture does not make the 1.873\,d period unlikely. The light curves are shown in Fig.\ref{o36}.\\

\subsection{GSC\,0460400743 (No 38)}
\begin{figure}[ht]
\centering
\includegraphics[width=0.45\textwidth,height=0.275\textheight]{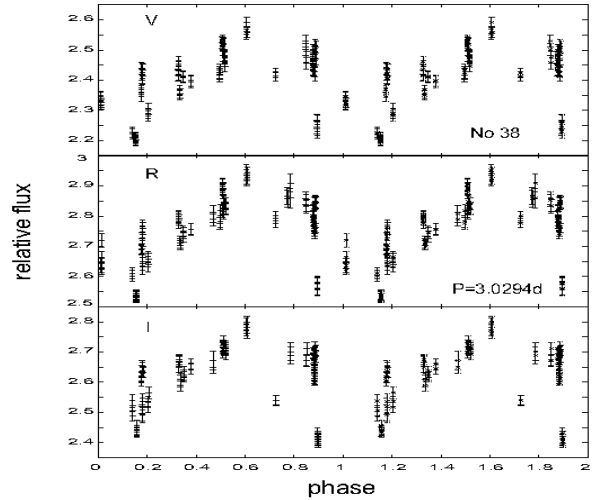}
\caption{The light curves of object 38. The folding was done with the period $P=3.0294$\,d.}
\label{o38}
\end{figure}

The approximately 8\,Myr old object GSC\,0460400743 is a K2 star and had been observed in 26 nights. Out of 1037 detected objects, about 30 stars were used to create the artificial CS.\\
\indent The period search was inconclusive between all three algorithms. For each filter and each algorithm different periods were found. For the $I$ filter \textit{Period04} and Lomb-Scargle gave a period of P=1.905\,d with a FAP of $10^{-12}$, the string\-length algorithm gave a period of 3.025\,d. For $R$ and $V$ filter, the fourier transform and Lomb-Scargle gave a period of 3.028\,d with a FAP of $10^{-17}$. The visual inspection excluded the 1.905\,d period. The obtained period is $P=3.029\pm0.003$\,d (Fig. \ref{o38}).

\subsection{GSC\,0460502733 (No 40)}
\begin{figure}[ht]
\centering
\includegraphics[width=0.45\textwidth,height=0.275\textheight]{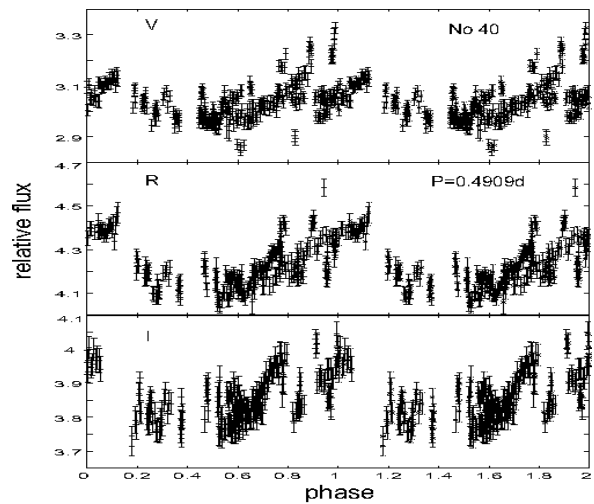}
\caption{The light curves of object 40. The folding was done with the period $P=0.4909$\,d.}
\label{o40}
\end{figure}

GSC\,0460502733 had been observed within 36 nights, 30 of them were used for the analysis. It is a K0 star with an estimated age of 6\,Myr. To create the artificial CS 15 stars were used.\\
The obtained period is 0.4909\,d. This period seems to be unlikely because it is almost half a siderial day, although the FAP is very low ($10^{-30}$). The phasefolding also does not fit very well (see Fig. \ref{o40}).
It is possible that the observed flux variations are due to irregular variability. To verifiy this, additional observations have to be made.

\subsection{GSC\,0460500170 (No 44)}
\begin{figure}[ht]
\centering
\includegraphics[width=0.45\textwidth,height=0.275\textheight]{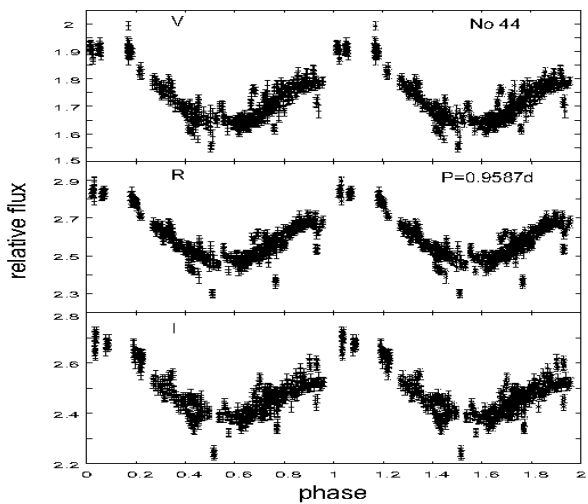}
\caption{Light curves of object 44 folded with a period of $P=0.9587$\,d.}
\label{o44}
\end{figure}

The object GSC\,0460500170 is a 2\,Myr old K4 star. The star had been observed within 27 nights, six of them were used for long term observations up to several hours. Because of the fact, that it was observed frequently and also in a series, it was possible to observe nearly the whole phase.\\
\indent The Lomb-Scargle algorithm and the fourier transform by \textit{Period04} gave similar results with a period of P=0.958\,d. With a FAP of $10^{-168}$ in the I filter and $10^{-189}$ in V filter this period seems to be determined very well.\\The stringlength algorithm gives as result a multiple of the obtained period. A phasefolding excludes this multiple, but confirms P=0.958\,d (Fig.\ref{o44}).

\subsection{GSC\,0447900348 (No 45)}
\begin{figure}[ht]
\centering
\includegraphics[width=0.45\textwidth,height=0.275\textheight]{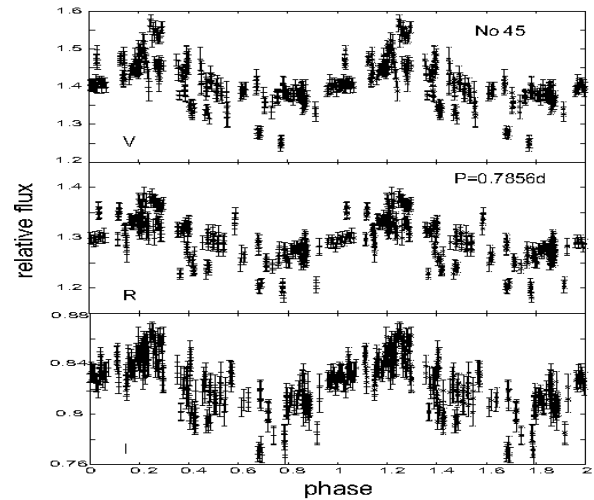}
\caption{The light curves of object 45 folded with a period of $P=0.7856$\,d.}
\label{o45}
\end{figure}

This star is a K2 star with an age estimated of 25\,Myr. Beside object 28 it is the oldest object within the sample.\\ \indent The object 45 had been observed within 36 nights, 34 were used for the analysis. Within the field 1708 objects were detected, only 20 were used to create the artificial CS.\\
The period search algorithms \textit{Period04} and Lomb-Scargle gave an identical period in all three filters of $P=0.7856$\,d (see Fig. \ref{o45}). The FAP obtained with the GLS is $10^{-52}$ in $I$ and $10^{-41}$ in $V$ filter. The stringlength algorithm calculated a period of 3.9294\,d which is the fivefold of the obtained 0.7856\,d period, in $I$ filter it calculated a period of 2.3572\,d.\\
\indent Because of the short period, dense time-series observations seem to be necessary to make sure, that the scattering of datapoints is due to observational and weather conditions.

\subsection{GSC\,0461001318 (No 46)}
The object with the designation GSC\,0461001318 is a K1 star with an estimated age of 6\,Myr. This object has been observed between 2007 August 1 and 2008 May 4 within 21 nights, 18 nights were used for light curve analysis.\\
\indent All algorithms give a similar result, an average\footnote{The average was calculated of the received values for all three filters} period of $P=5.735\pm0.003$\,d. The FAP given by the GLS is $10^{-18}$. 
\begin{figure}[ht]
\centering
\includegraphics[width=0.45\textwidth,height=0.275\textheight]{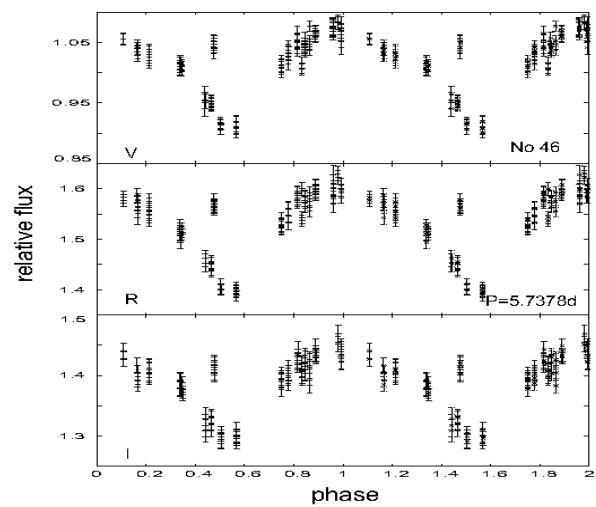}
\caption{The light curves of object 46 folded with a period of $P=5.7378$\,d.}
\label{o46}
\end{figure}

\section{Summary and Conclusions}
\begin{figure}[ht]
\centering
\includegraphics[width=0.48\textwidth]{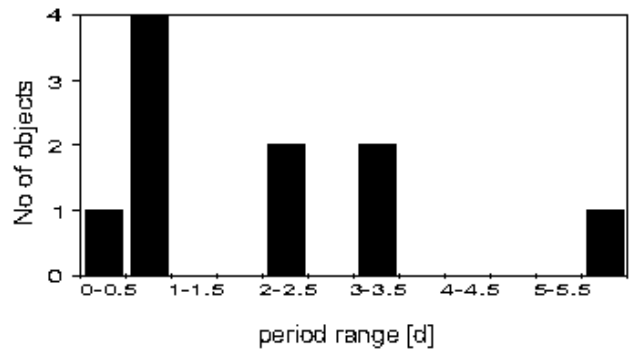}
\caption{Histogramm of all observed periods. The maximum lies between 0.5 and 1\,d period.}
\label{bar}
\end{figure}

At the University Observatory Jena we observed ten objects within the SFR in Cep-Cas in order to determine rotational periods of wTTS. The periods vary between 0.49\,d and 5.7\,d. Five objects within the sample show periods less than one day, the minimum observed period is 0.4909\,d, whereas this period seems to be unlikely; the maximum period within this period range is 0.9587\,d. Four objects show periods between 2 and 4 days, and only one object (No 46) has a period $>5$\,d. Table \ref{GSHres} gives an overview of the results of the period search and the final period used for the phasefolding.  The periods of the objects 44 (P=0.9587\,d) and object 46 (P=5.7378\,d) are the best and most significant ones. The periods of target 19, 20, 40 and 45 have to be confirmed.\\
\indent Fig.\ref{bar} shows the histogramm of all observed periods.
\begin{table}[ht]
\centering
\caption {Results of the period search with Period04 (P04), Lomb-Scargle (LS) and Stringlength (SL) and the final period for the observed objects. The errors consider the results of the period search averaged over similar results of the period search algorithms.}
\label{GSHres}
\begin{tabular}{cccccc}\hline
object & filter & P04  & LS & SL & final \\
No$^a$ & & P [d] & P [d] & P [d] & P [d]\\ \hline
19 & $V$ & 1.02 & 1.46 & 4.34 & \\
 & $R$ & 1.08 & 1.08 & 2.83 & 1.08 or \\
 & $I$ & 1.08 & 0.82 & 3.89 & 0.82 \\ \hline
20 & $V$ & 2.54 & 1.56 & 2.54 & \\
 & $R$ & 2.54 & 0.88 & 2.66 &  \\
 & $I$ & 0.88 & 0.88 & 4.78 & 0.886$\pm$0.002 \\ \hline
28 & $V$ & 2.11 & 2.11 & 2.11 & \\
 & $R$ & 2.11 & 2.11 & 1.52 &  \\
 & $I$ & 2.14 & 2.14 & 1.75 & 2.13$\pm$0.01 \\ \hline
34 & $V$ & 3.20 & 3.20 & 1.69 & \\
 & $R$ & 1.07 & 1.07 & 1.07 &  \\
 & $I$ & 3.21 & 1.04 & 1.07 & 3.203$\pm$0.001 \\ \hline
36c1 & $V$ & 2.03 & 2.04 & 1.87 & \\
 & $R$ & 2.13 & 2.03 & 2.13 &  \\
 & $I$ & 2.03 & 2.13 & 1.87 & 2.136$\pm$0.004 \\ \hline
38 & $V$ & 3.03 & 3.03 & 3.03 & \\
 & $R$ & 3.03 & 3.03 & 3.64 &  \\
 & $I$ & 1.90 & 1.90 & 3.03 & 3.029$\pm$0.003 \\ \hline
40 & $V$ & 0.49 & 0.95 & 1.92 & \\
 & $R$ & 0.49 & 0.49 & 2.74 &  \\
 & $I$ & 0.30 & 0.49 & 2.36 & 0.491 (?) \\ \hline
44 & $V$ & 0.96 & 0.96 & 3.83 & \\
 & $R$ & 0.96 & 0.96 & 3.83 &  \\
 & $I$ & 0.96 & 0.96 & 3.83 & 0.959$\pm$0.001 \\ \hline
45 & $V$ & 0.79 & 0.79 & 3.66 & \\
 & $R$ & 0.79 & 0.79 & 3.93 &  \\
 & $I$ & 0.79 & 0.79 & 2.36 & 0.786 \\ \hline
46 & $V$ & 5.73 & 5.73 & 5.73 & \\
 & $R$ & 5.74 & 5.74 & 5.73 &  \\
 & $I$ & 5.73 & 5.73 & 5.73 & 5.738$\pm$0.002 \\ \hline
\end{tabular}
\\$^a$ Number of object taken from Tachihara et al. (2005)
\end{table}
To determine the periods the best algorithm seems to be the Lomb-Scargle, although stringlength works without fitting sine waves. In case of short periodic variations it reveals its shortcoming by determining multiples of the correct period. To improve this the sampling has to be adjusted. The fourier transformation by \textit{Period04} works similar to Lomb-Scargle and obtains for large data sets nearly the same results. Advantage of Lomb-Scargle is the implemented least-square fitting, which has to be carried out as an additional step in the \textit{Period04} algorithm.
The amplitudes of magnitude variation range between 0.1 mag in $I$ and 0.3 mag in $V$. The mean photometric accuracy depends on the brightness of the object. The faintest object No 34 ($V=13.09$\,mag) has an accuracy of 0.01\,mag, the precision of the brightest one (No 19, $V=10.39$\,mag) is 0.007\,mag. The regular periodic variability amplitudes of typically $\pm0.2$\,mag in our observations are typical for dark, cold spots.\\
\indent The nine periods between 0.5 and 4 days are in good agreement with results of rotation period distributions for low mass stars ($0.4<M<1.5M_\odot$) described e.g. in Bouvier et al.\,(1997) or Herbst et. al\,(2007). The observation of rotational periods of PMS and zero age main sequence (ZAMS) stars help to understand the evolution of angular momentum.
\\[12pt]
\textit{Acknowledgements.} The authors would like to thank the \\GSH Observer Team for the nightly observations. Moreover we thank the technical staff of the University Observatory Jena in Gro\ss{}schwabhausen, especially Tobias B\"ohm. AK acknowledges support from the German National Science Foundation (Deutsche Forschungsgemeinschaft, DFG) in grant KR 2164/8-1, RN acknowledges general support from DFG in grants NE 515/13-1, 13-2, and 23-1, TE and MMH acknowledge support from the DFG in SFB/TR 7 Gravitational Wave Astronomy, TOBS acknowledges support from the Evangelisches Studienwerk e.V. Villigst, SR and MV acknowledge support from the EU in the FP6 MC ToK project MTKD-CT-2006-042514, TR acknowledges\\support from DFG in grant NE 515/23-1 and MMoualla acknowledges support from the government of Syria.

\newpage
\appendix
\section{Output of period search algorithms}
\begin{figure}[ht]
\centering
\includegraphics[width=0.48\textwidth]{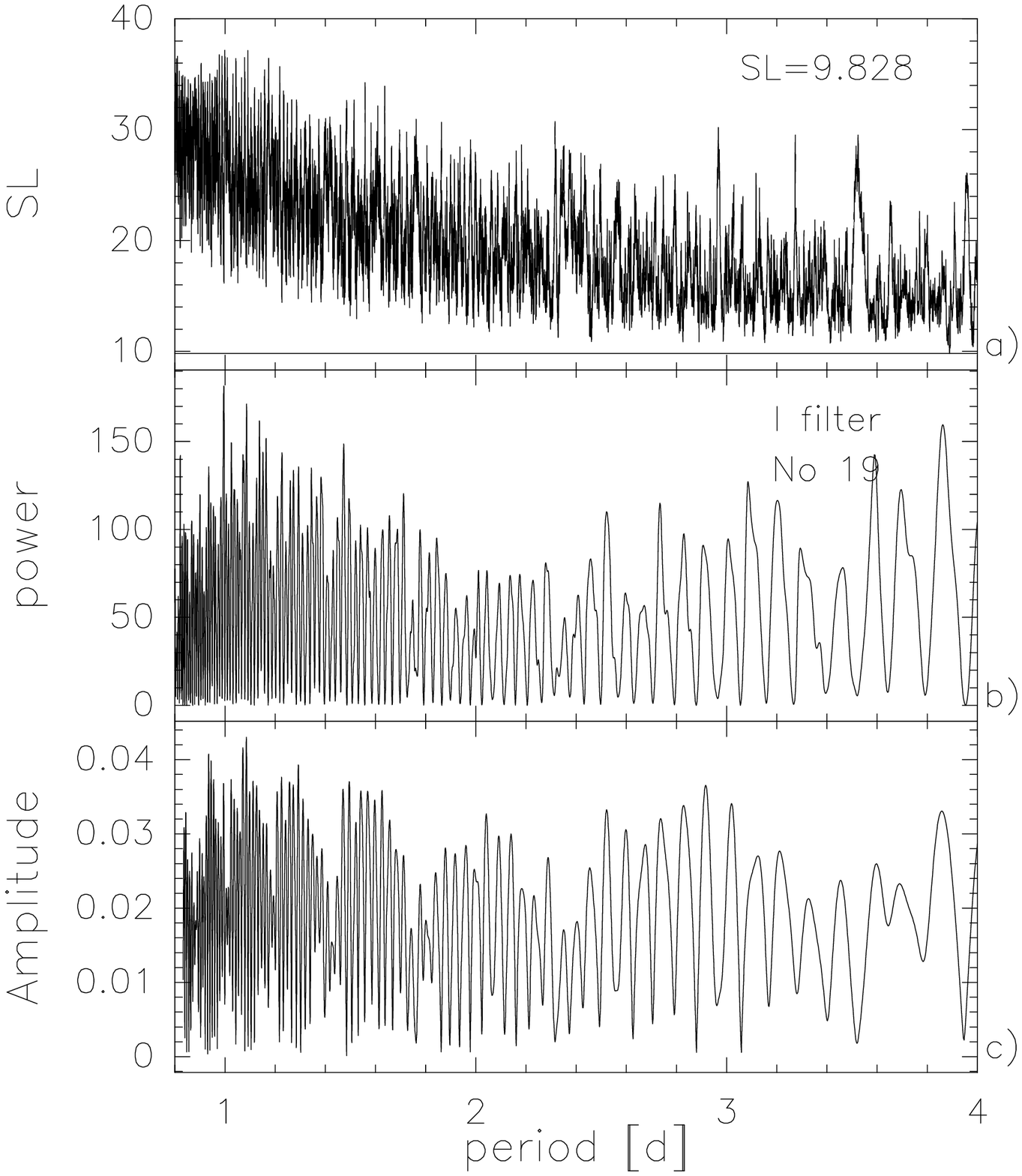}
\caption{Output for object No 19: a) Stringlength algorithm, b) generalised Lomb-Scargle and c) the Fourier transform given by \textit{Period04}.}
\label{o19_period}
\end{figure}

\begin{figure}[ht]
\centering
\includegraphics[width=0.48\textwidth]{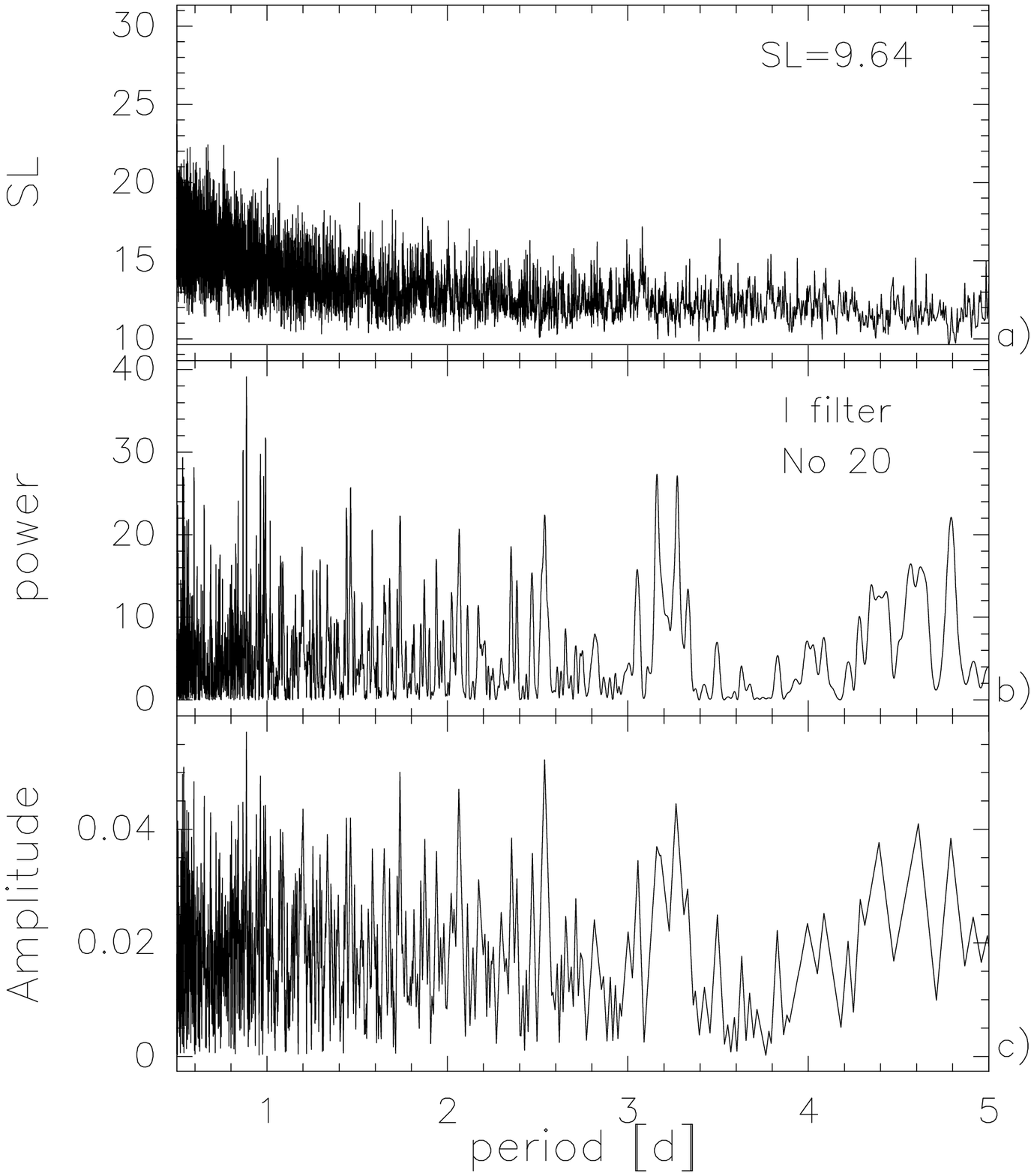}
\caption{Output for object No 20: a) Stringlength algorithm, b) generalised Lomb-Scargle algorithm and c) the Fourier transform given by \textit{Period04}.}
\label{o20_period}
\end{figure}

\begin{figure}[ht]
\centering
\includegraphics[width=0.48\textwidth]{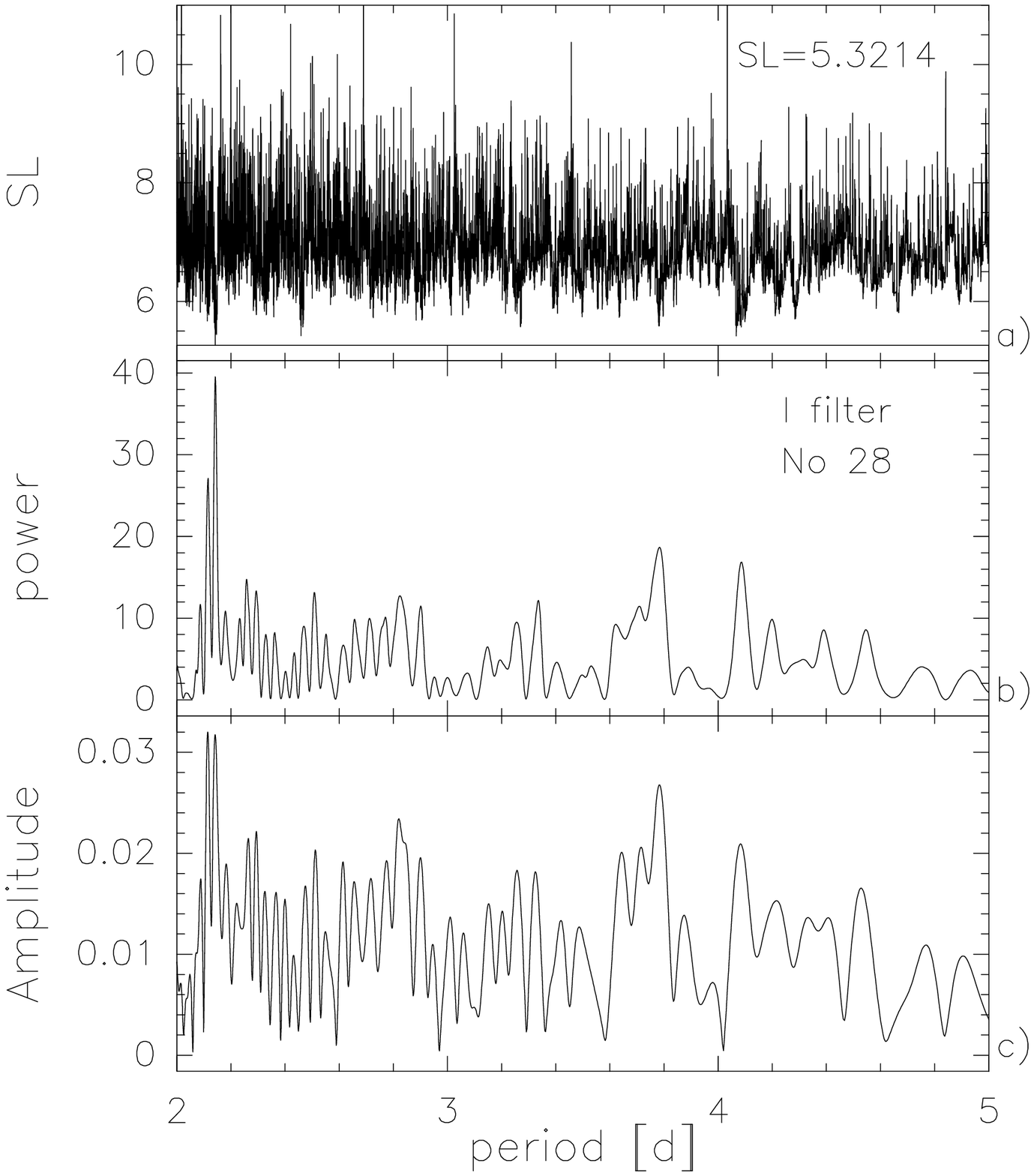}
\caption{Output for object No 28: a) Stringlength algorithm, b) generalised Lomb-Scargle and the c) the Fourier transform given by \textit{Period04}.}
\label{o28_period}
\end{figure}

\begin{figure}[ht]
\centering
\includegraphics[width=0.48\textwidth]{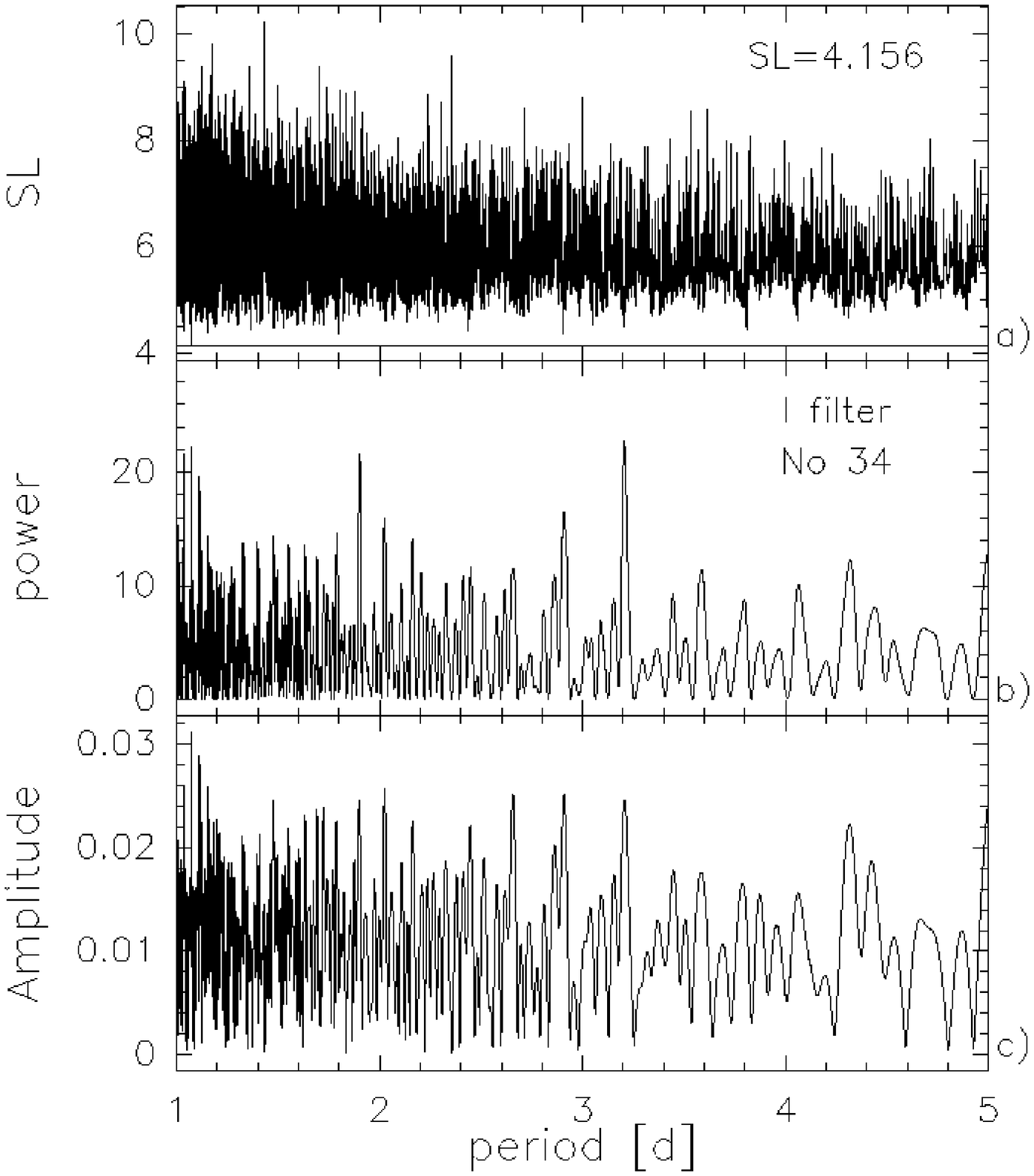}
\caption{Output for object No 34: a) Stringlength algorithm, b) generalised Lomb-Scargle and c) the Fourier transform given by \textit{Period04}.}
\label{o34_period}
\end{figure}

\begin{figure}[ht]
\centering
\includegraphics[width=0.48\textwidth]{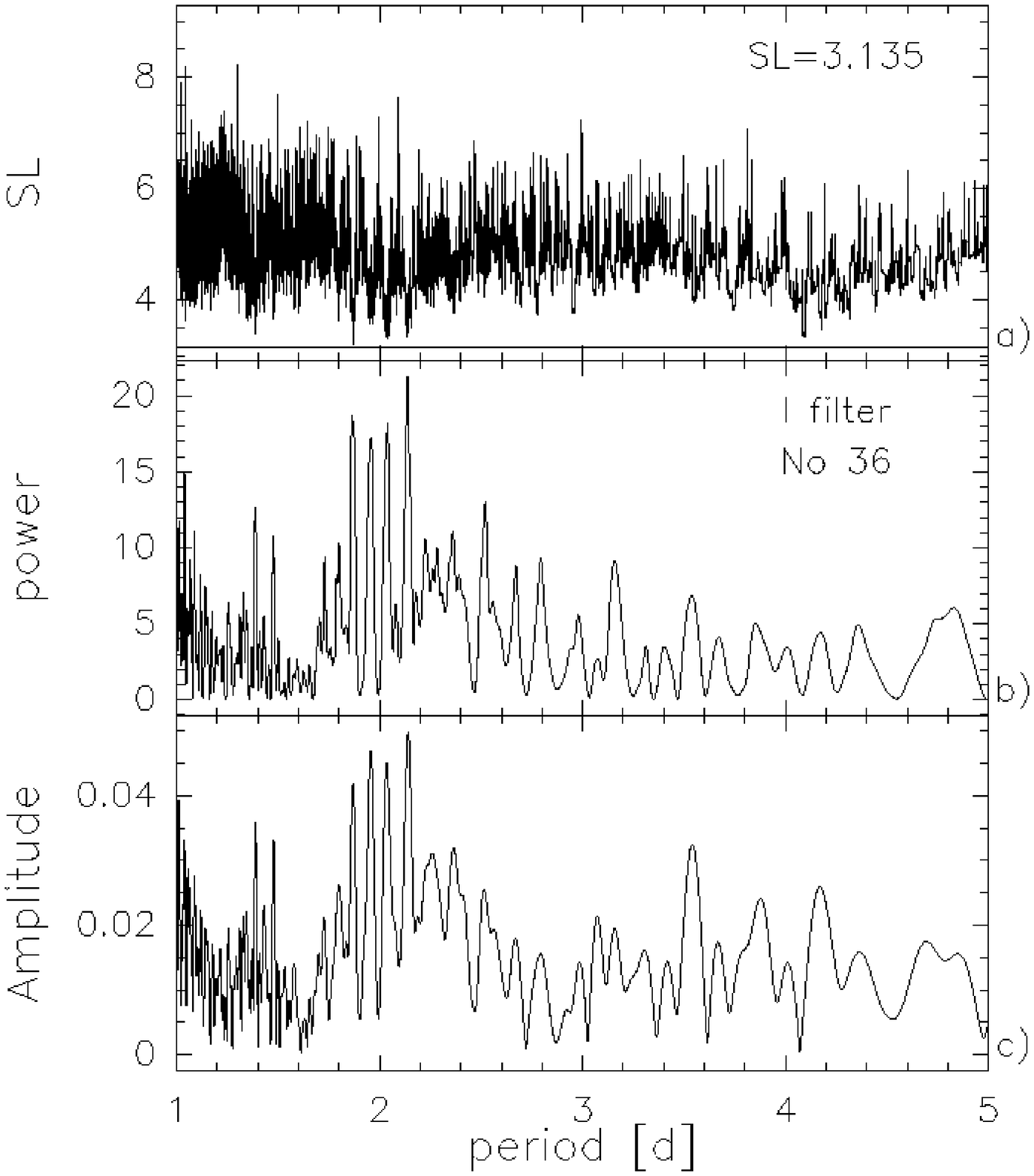}
\caption{Output for object No 36: a) Stringlength algorithm, b) the generalised Lomb-Scargle periodogram analysis and c) the Fourier transform given by \textit{Period04}.}
\label{o36_period}
\end{figure}

\begin{figure}[ht]
\centering
\includegraphics[width=0.48\textwidth]{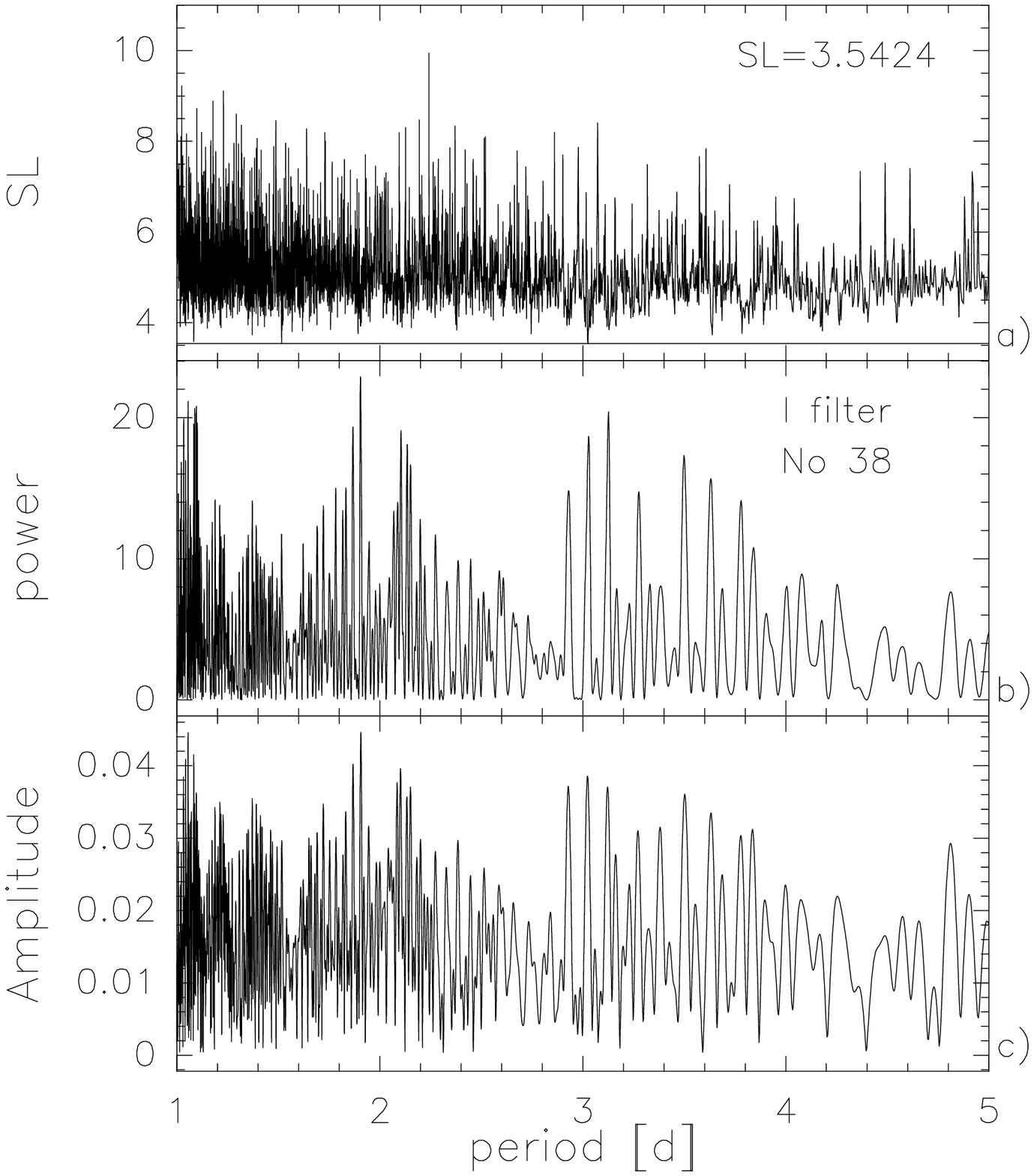}
\caption{Output for object No 38: a) Stringlength algorithm, b) the generalised Lomb-Scargle periodogram analysis and c) the Fourier transform given by \textit{Period04}.}
\label{o38_period}
\end{figure}

\begin{figure}[ht]
\centering
\includegraphics[width=0.48\textwidth]{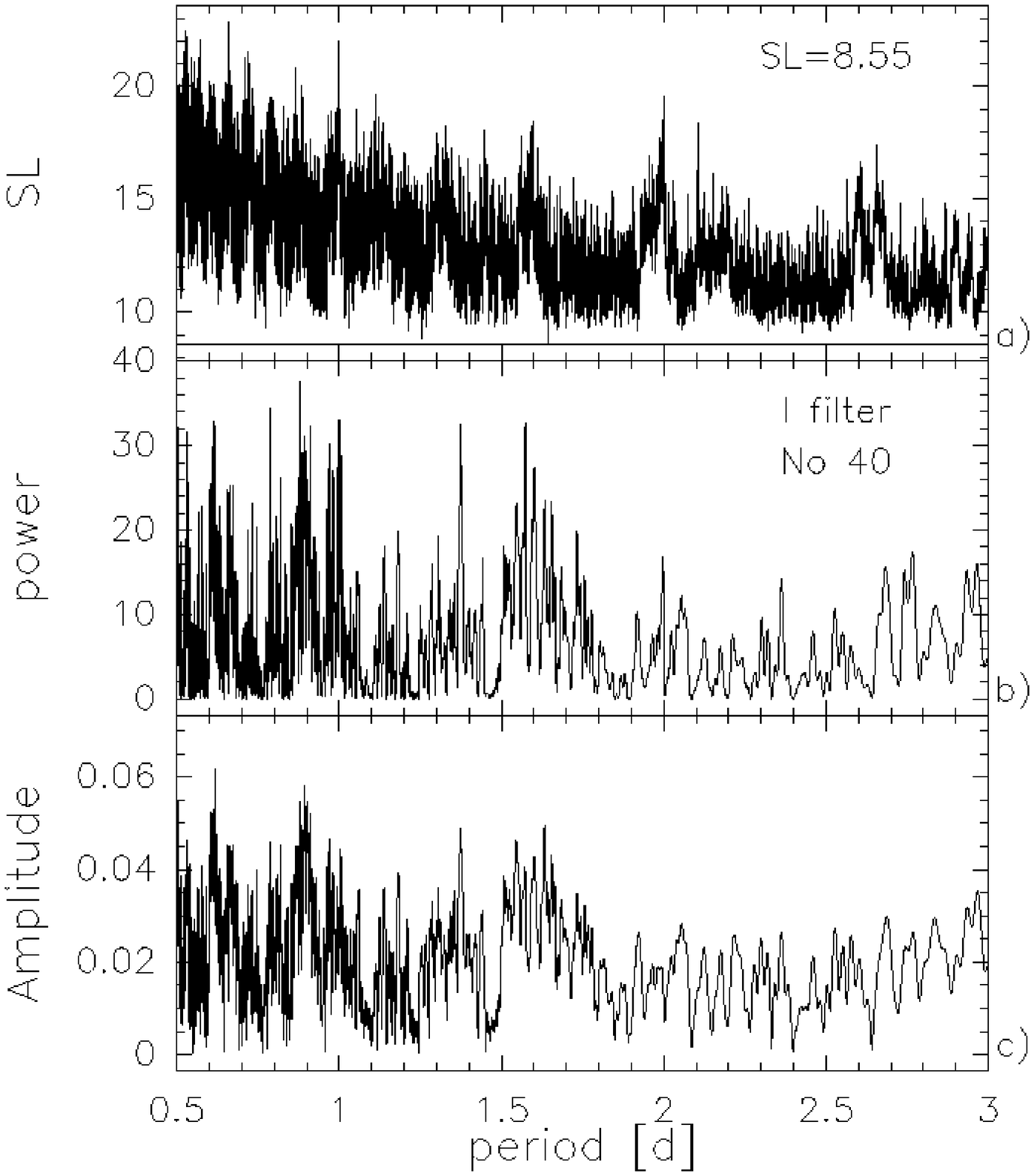}
\caption{Output for \textit{object No 40}: a) Stringlength algorithm, b) the generalised Lomb-Scargle periodogram analysis and c) the fourier transform by \textit{Period04}.}
\label{o40_period}
\end{figure}

\begin{figure}[ht]
\centering
\includegraphics[width=0.48\textwidth]{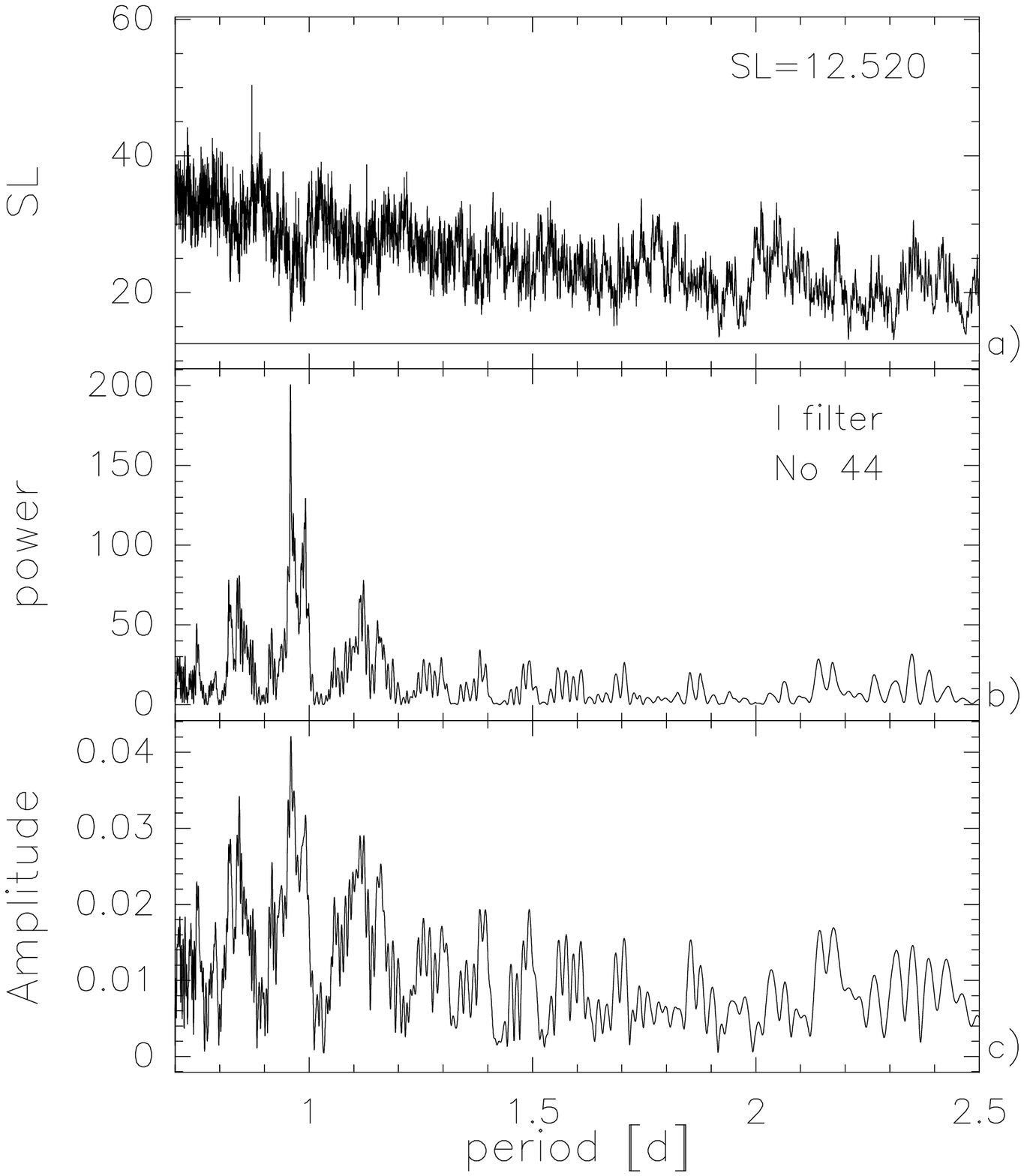}
\caption{Output for object No 44: a) Stringlength algorithm, b) generalised Lomb-Scargle algorithm and c) the Fourier transform given by \textit{Period04}.}
\label{o44_period}
\end{figure}

\begin{figure}[ht]
\centering
\includegraphics[width=0.48\textwidth]{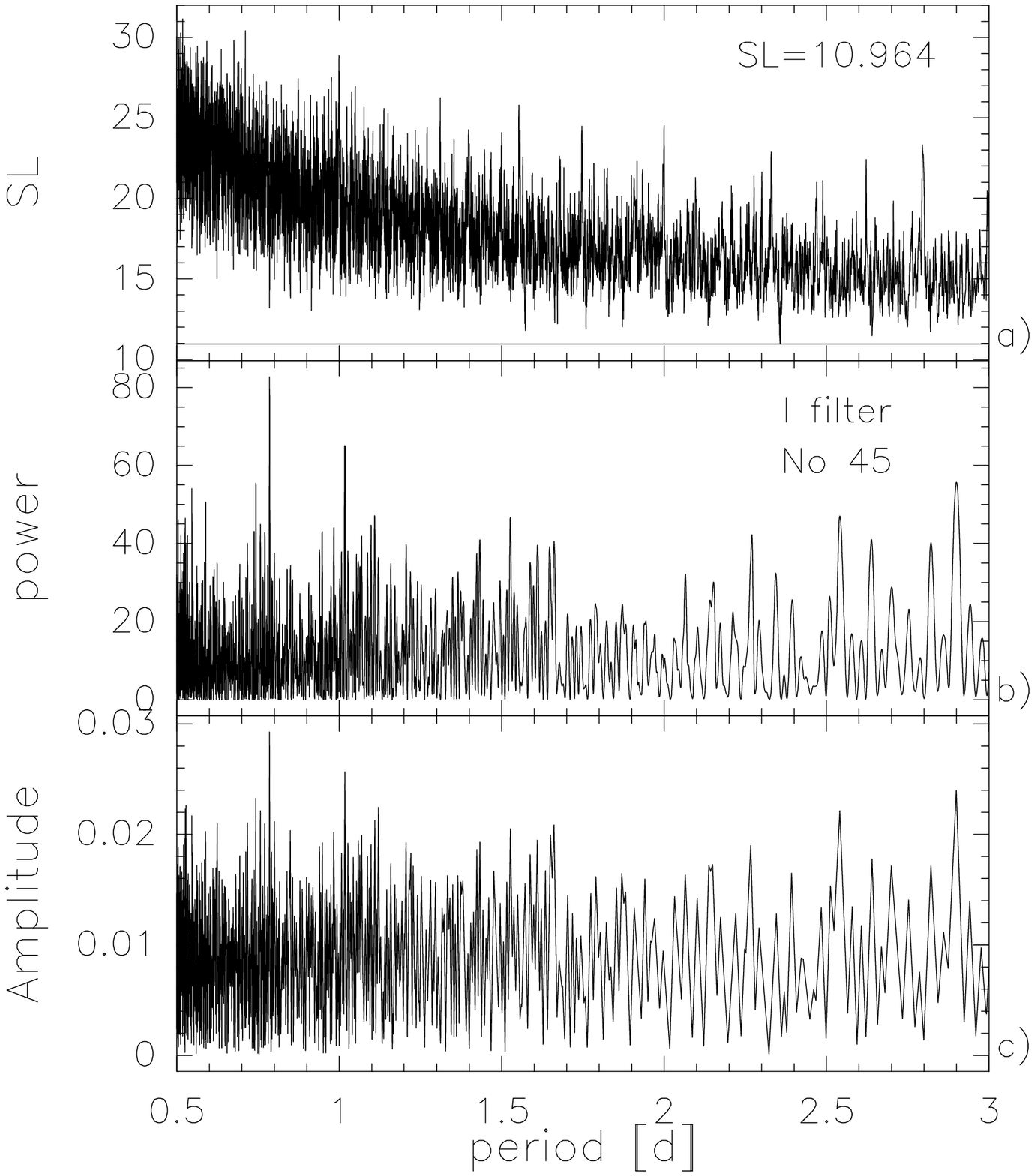}
\caption{Output for object No 45: a) Stringlength algorithm, b) generalised Lomb-Scargle periodogram analysis and c) the Fourier transform given by \textit{Period04}.}
\label{o45_period}
\end{figure}

\begin{figure}[ht]
\centering
\includegraphics[width=0.48\textwidth]{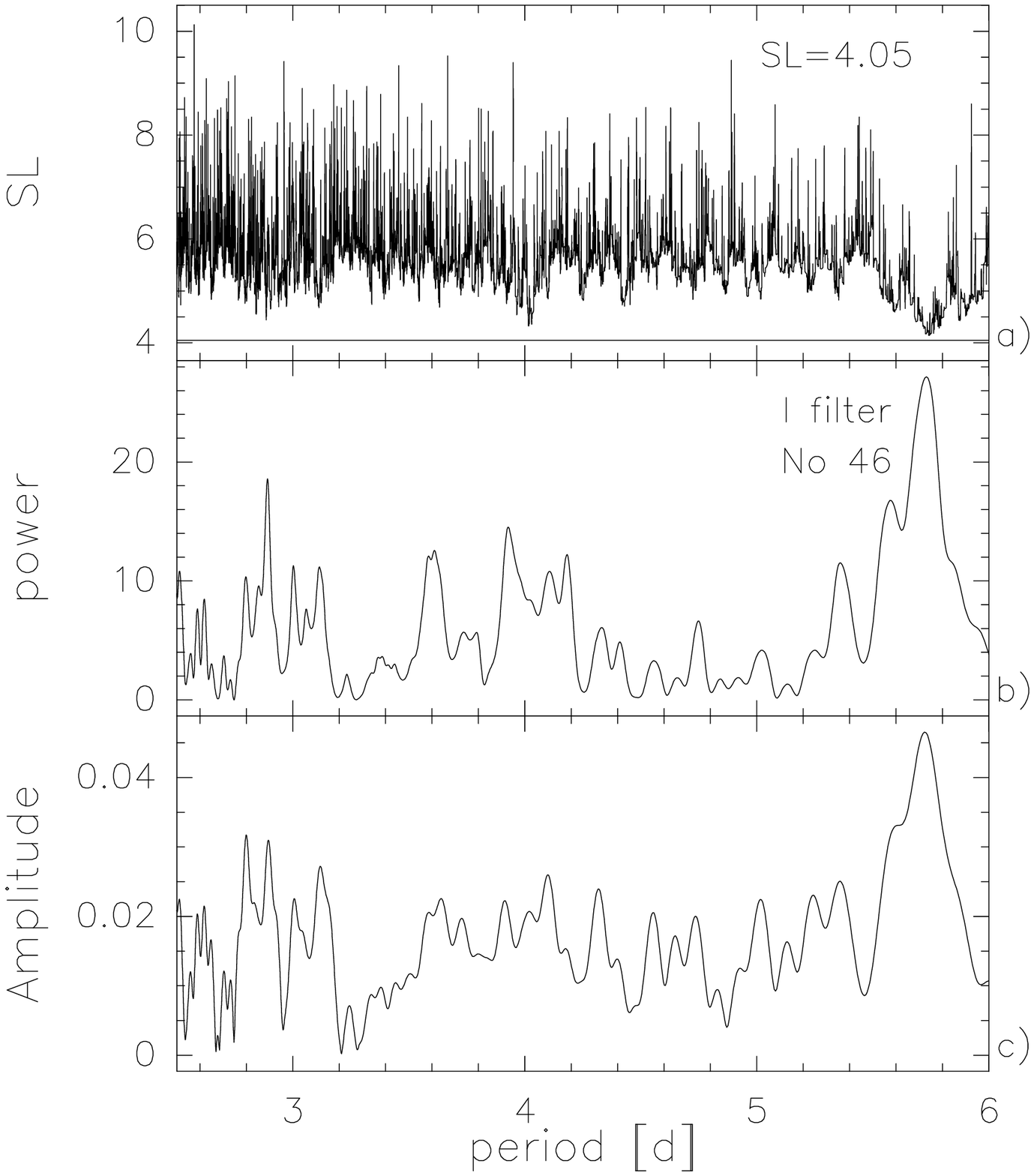}
\caption{Output for object No 46: a) Stringlength algorithm, b) generalised Lomb-Scargle periodogram analysis and c) the Fourier transform given by \textit{Period04}.}
\label{o46_period}
\end{figure}

In the figures \ref{o19_period} to \ref{o46_period} the output of the different algorithms are shown. The stringlength value marked with a horizontal line in the figures \ref{o19_period} a) to \ref{o46_period} a) is the shortest stringlength that marks the period obtained with the String\-length algorithm, given in Table \ref{GSHres}.\\
For each periodogram the a) stringlength, b) power and c) amplitude is plotted over the period P.
\end{document}